\documentclass[preprint,nofootinbib,a4paper]{revtex4}
\usepackage{amsmath}
\usepackage{amsfonts}

\begin{document}

\title[Universal properties of distorted Kerr-Newman black holes]
{Universal properties of distorted\\  Kerr-Newman black holes}

\author{Marcus Ansorg}
\affiliation{Institute of Biomathematics and Biometry,
 Helmholtz Zentrum M\"unchen,
 Ingolst\"adter Landstr. 1,
 D-85764 Neuherberg, Germany}
\email{marcus.ansorg@helmholtz-muenchen.de}
\author{J\"org Hennig}
\affiliation{Max Planck Institute for Gravitational Physics,
Am M\"uhlenberg 1, D-14476 Golm, Germany}
\email{Joerg.Hennig@aei.mpg.de, Carla.Cederbaum@aei.mpg.de}
\author{Carla Cederbaum}
\affiliation{Max Planck Institute for Gravitational Physics,
Am M\"uhlenberg 1, D-14476 Golm, Germany}

\begin{abstract}
We discuss universal properties of axisymmetric and stationary
configurations consisting of a central black hole and surrounding matter in
Einstein-Maxwell theory. In particular, we find that certain physical equations
and inequalities (involving angular momentum, electric charge and horizon area)
are not restricted to the Kerr-Newman solution but can be generalized
to the situation where the black hole is distorted
by an arbitrary axisymmetric and stationary surrounding matter distribution.
\end{abstract}

\maketitle

\section{Introduction}

One of the most important exact solutions in General Relativity is the
Kerr metric describing the spacetime around axisymmetric and stationary
black holes in vacuum. This solution is particularly
interesting because it is generally expected that it will be
approached asymptotically by any collapsing matter distribution --
provided that all the matter finally falls into the black hole.
It is, however, conceivable that for a long time period 
a substantial portion of the matter distribution moves around the 
black hole in a quasi-stationary state. Such a configuration can be approximated well
by an equilibrium system consisting of a central
black hole and surrounding matter, e.g., a fluid ring or disk
(describing a black hole with an accretion disk or a galaxy with a
central black hole). 

Little is known about how such
surrounding matter influences the central black hole. 
In this essay we discuss several
\emph{universal} properties that black holes possess independent of
the nature of their environment. Our investigations apply to the 
more general case in which additional electromagnetic
fields are considered, i.e. we  study ``distorted Kerr-Newman black
holes'' within Einstein-Maxwell theory.

We point out that for stellar black holes it is not expected that surrounding  matter contributes more then a few percent to the total mass of the system. Hence, for such configurations the influence of the matter on the black hole should be small or even negligible. On the other hand, for a galactic black hole the surrounding matter (the galaxy) contributes the main part of the mass and dominates the bahaviour of the spacetime in larger distances. However, in a vicinity of a galactic black hole, it is expected that the deviation of the spacetime geometry from a Kerr metric is again very small. But even though astrophysically relevant black holes might typically be close to the Kerr solution, the following universal properties are of fundamental theoretical interest since they provide \emph{exact} relations (equations and inequalities) which are completely independent of the amount and type of matter surrounding a black hole.

\section{A universal inequality for sub-extremal black holes}

It is well-known that the three-parameter Kerr-Newman family of solutions
describes single rotating, electrically charged, axisymmetric and stationary black holes in
electrovacuum. For the parameters one can choose, e.g., the mass $M$, the angular momentum $J$, and the electric charge
$Q$ of the black hole\footnote{In the special case of the Kerr-Newman
metric, these quantities can be read off from the asymptotic behavior of the metric coefficients.}. 
For a solitary black hole, which is characterized by the existence of an event horizon, the relation
\begin{equation}\label{par}
 \left(\frac{J}{M}\right)^2+Q^2\le M^2
\end{equation}
must be satisfied. Equivalently,
expressing $M$ through $J$, $Q$, and the area $A$ of the event horizon,
we find that the three parameters $J$, $Q$, and $A$ have to obey
\begin{equation}\label{ineq}
 (8\pi J)^2+(4\pi Q^2)^2\le A^2\,.
\end{equation}
Now, a Kerr-Newman black hole with equality in \eqref{par} (or \eqref{ineq}) is called \emph{extremal}, otherwise  \emph{sub-extremal}. 

A different definition of sub-extremality of black holes refers to the existence of trapped surfaces in every sufficiently
small interior neighborhood of their event horizons \cite{Booth}. In the Kerr-Newman family, both notions of sub-extremality turn out to be equivalent. However, the concept of trapped surfaces is more general and applies to arbitrary axisymmetric and stationary black holes. Hence it is appropriate to use the corresponding definition of sub-extremality, and we will do so in the following.

In \cite{Hennig2008, Hennig2010}, we have investigated such sub-extremal stationary and
axisymmetric black holes surrounded by matter in full Einstein-Maxwell
theory and found that inequality \eqref{ineq} strictly holds:
\begin{equation}\label{sineq}
 (8\pi J)^2+(4\pi Q^2)^2< A^2.
\end{equation}
Through this result we generalize the link between the two notions of sub-extremality to the realm of 
distorted black holes, i.e. we establish a universal relation between the
geometric concept of existence of trapped surfaces and the inequality to
be obeyed by the physical black hole parameters $J$, $Q$, and $A$.
\footnote{Note that these quantities can be determined locally on the event horizon.
We further remark that the definition of
$J$ (in contrast to that of $Q$ and $A$) involves a choice: The total
angular momentum of the 
spacetime is composed of matter, electromagnetic field, and black hole
contributions. While clearly the matter part should be excluded from the
definition of $J$, both a Komar integral and an appropriate
electromagnetic event horizon integral is taken into account for our
notion of the black hole's local angular momentum (cf. \cite{AnsorgPfister} 
for a more thorough discussion).}

For the proof of (\ref{sineq}) for sub-extremal distorted black holes 
it suffices to study the Einstein-Maxwell equations in an
electrovacuum vicinity of the event horizon. Using these equations,
it is possible to reformulate \eqref{sineq} in terms of a
variational problem: An appropriate functional depending on the
horizon values of three metric and electromagnetic potentials must
always be greater than or equal to $1$. As shown in \cite{Hennig2008,
Hennig2010}, this variational problem can be analyzed with methods from
the calculus of variations.

A fascinating application of (\ref{sineq}) is the
solution of an old problem in General Relativity, the \emph{balance
problem} for two black holes: Is it possible that two aligned
axisymmetric and
stationary black holes are in equilibrium, i.e. can the spin-spin
repulsion of two rotating black holes compensate their gravitational
attraction? The application of a particular soliton method (the {\em
inverse scattering method})\footnote{\label{foot:Soliton}Soliton methods 
are based on the existence of a \emph{linear} matrix problem
which is equivalent to the \emph{nonlinear} field equations via its integrability condition
\cite{Neugebauer1983}. (See \cite{Neugebauer1996} for a sophisticated
introduction to soliton methods in General Relativity.)},
provides a proof of the non-existence of such equilibrium situations, 
showing that at least one of the two black holes in this ensemble would violate inequality
\eqref{sineq}, cf. \cite{Neugebauer2009}.

\section{The interior of the black hole}

Having discussed universal properties on a black hole's event horizon,
we will now direct our attention to its interior. Again, we were able
to generalize a well-known feature of Kerr-Newman solutions, namely the
existence of a second horizon inside the black hole  --
the \emph{inner Cauchy horizon} -- which is defined as
the future boundary of the
domain of dependence of the event horizon. Since predictability breaks
down beyond the Cauchy horizon, its stability is an
important issue. Based on an argument by R.~Penrose \cite{Penrose1968},
it was assumed that generic perturbations grow infinitely near
the Cauchy horizon. However, for purely axisymmetric and
stationary (arbitrarily large) perturbations, i.e. for our situation of 
a central black hole
with surrounding matter, a regular inner Cauchy horizon can be shown 
to exist provided the angular momentum $J$ and the charge $Q$ of
the black hole do not vanish simultaneously. Moreover, the region
between event horizon and Cauchy horizon is then completely
regular.\footnote{A particular exact solution for the interior black
hole region is discussed in \cite{Abdolrahimi}.}
In contrast, the Cauchy horizon becomes singular and
approaches a scalar curvature singularity in
the limit $J\to 0$, $Q\to 0$ (precisely as in the transition
from a Kerr-Newmann black hole to a Schwarzschild black
hole).

Remarkably, if the inner Cauchy horizon exists (i.e.
if $J$ and $Q$ do not vanish simultaneously), 
then the area $A_\textrm{CH}$ of the Cauchy horizon 
and the event horizon area $A$ are related via
\begin{equation}\label{rel}
 (8\pi J)^2+(4\pi Q^2)^2=A_\textrm{CH} A.
\end{equation}
Note that the left hand side of \eqref{rel} is the same as in \eqref{sineq}.

The proof of these statements \cite{Ansorg2008,Ansorg2009,Hennig2009}
utilizes again soliton methods. From an analysis of the linear matrix problem (see footnote \ref{foot:Soliton}) 
on a closed path (along the event horizon, the Cauchy
horizon and the two parts of the symmetry axis, which connect both horizons), 
it is possible to deduce the existence of a Cauchy horizon. Moreover,
explicit formulas for the metric and electromagnetic potentials on the
Cauchy horizon emerge, and these are used to show
\eqref{rel}. The second essential ingredient of the proof is a
theorem by P.~Chru\'sciel \cite{Chrusciel1990}.

Finally, we note that the interior spacetime region of axisymmetric and
stationary black holes is closely related to Gowdy spacetimes
(cosmological models with two spacelike Killing fields), making
the above statements applicable in this cosmological context as well. 
They also imply the analogous existence and regularity statements for Cauchy horizons~\cite{Hennig2010b}.
\section{Configurations with degenerate black holes}

In the degenerate limit of converging Cauchy and event horizons (expressed by vanishing surface gravity $\kappa$), we obtain equality in equation \eqref{ineq}, cf.~\cite{AnsorgPfister}:
\begin{equation}\label{deg}
 (8\pi J)^2+(4\pi Q^2)^2=A^2.
\end{equation}
Consistently, equation \eqref{deg} also follows
from equation \eqref{rel} as the areas $A$ and
$A_\textrm{CH}$ coincide in the degenerate limit.

As a consequence of \eqref{deg}, the characterizing relation
\begin{equation}\label{deg2}
 \left(\frac{J}{M}\right)^2+Q^2=M_{CR}^2
\end{equation}
of the extremal Kerr-Newman solution continues to hold in 
presence of surrounding matter -- in accordance with the fact that Kerr-Newman black holes are degenerate precisely if they are extremal. (In this general situation, $M_{CR}$ denotes the
Christodoulou-Ruffini mass \cite{Christodoulou} of the black hole which agrees with the ADM mass in the Kerr-Newman setting.)

\section{Discussion}

We have demonstrated that the following statements about the Kerr-Newman
black hole are more generally valid for any axisymmetric and stationary
black hole with surrounding matter in Einstein-Maxwell theory:
\begin{itemize}
 \item Physically relevant black holes are expected to be sub-extremal,
       where sub-extremality is defined by the existence of trapped surfaces
       in every sufficiently small interior neighborhood of the event horizon. 
       Angular momentum $J$ and electric charge $Q$
       of a sub-extremal black hole are bounded in terms of
       its horizon area $A$: $(8\pi J)^2+(4\pi Q^2)^2<A^2$.
 \item The black hole possesses a second horizon -- an inner Cauchy
       horizon -- if and only if $J$ and $Q$ do not vanish simultaneously.
       The Cauchy horizon is regular whenever it exists.
 \item The spacetime region between event horizon and Cauchy horizon is
       completely regular.
 \item The horizon areas satisfy the universal relation
       $(8\pi J)^2+(4\pi Q^2)^2=A_\textrm{CH} A$.
 \item The relations
       $(8\pi J)^2+(4\pi Q^2)^2=A^2$ and
       $\left(\frac{J}{M}\right)^2+Q^2=M_{CR}^2$ hold for 
       degenerate black holes ($M_{CR}:$ 
       Christodoulou-Ruffini mass). In that sense, degenerate black
       holes could be named \emph{extremal}. 
\end{itemize}

It would be interesting to identify further universal properties of the
black hole configurations in question, e.g. black holes in Yang-Mills
theory or in higher dimensions. 

\begin{acknowledgments}
We would like to thank David Petroff for commenting on the manuscript. 
This work was supported by the Deutsche
For\-schungsgemeinschaft (DFG) through the
Collaborative Research Centre SFB/TR7
``Gravitational Wave Astronomy'' and by the International Max Planck
Research School (IMPRS) "Geometric Analysis and Gravitation".
\end{acknowledgments}

\end{document}